\documentclass[a4paper,USenglish]{lipics-v2018}

\usepackage{amsmath, amsfonts}
\usepackage{graphicx}
\usepackage[ruled,vlined]{algorithm2e}
\usepackage{lineno}
\nolinenumbers
\title{Mapper Comparison with Wasserstein Metrics}
\author{Michael McCabe}{Ayasdi, Menlo Park, CA}{michael.mccabe@ayasdi.com}{}{}
\hideLIPIcs

\subjclass{Theory of Computation $\rightarrow$ Computational Geometry, Mathematics of Computing $\rightarrow$ Graph Algorithms, Mathematics of Computing $\rightarrow$ Topology}

\authorrunning{M.\,McCabe}
\titlerunning{Mapper Comparison with Wasserstein Metrics}

\Copyright{Michael McCabe}

\begin{document}

\maketitle

\begin{abstract}The challenge of describing model drift is an open question in unsupervised learning. It can be difficult to evaluate at what point an unsupervised model has deviated beyond what would be expected from a different sample from the same population. This is particularly true for models without a probabilistic interpretation. One such family of techniques, Topological Data Analysis, and the Mapper algorithm in particular, has found use in a variety of fields, but describing model drift for Mapper graphs is an understudied area as even existing techniques for measuring distances between related constructs like graphs or simplicial complexes fail to account for the fact that Mapper graphs represent  a combination of topological, metric, and density information. In this paper, we develop an optimal transport based metric which we call the Network Augmented Wasserstein Distance for evaluating distances between Mapper graphs and demonstrate the value of the metric for model drift analysis by using the metric to transform the model drift problem into an anomaly detection problem over dynamic graphs.
\end{abstract}

\section{Introduction}
\label{intro}
An important consideration when using machine learning models in production settings is the question of when a model must be retrained. More often than not there is a cost associated with retraining models beyond just the computational expense. In the financial sector, regulators frequently require new justification to be constructed in order to deploy new models. In the population health world, intervention programs might be built around a specific patient segmentations and replacing that patient segmentation model could require a rework of the associated intervention programs as well. There are many reasons why re-deploying models can be non-trivial, and when it is, users want to maximize the lifespan of models while ensuring that decisions are not being built off of out-of-date information.

For supervised models, there are a number of practical approaches to address this issue. The simplest of these approaches are based on the idea that when model performance degrades, it is time to retrain. However, the majority of available data is unlabeled and can only be analyzed in an unsupervised setting. The question of when an unsupervised model no longer fits the data can be more complicated as there is no longer a clear measure of “correctness”. Generative models naturally provide a likelihood function that can be used to evaluate how probable it is that we'd see our new data given that our model is a good fit for the generative process \cite{zong2018deep,chandola2009anomaly}. Unfortunately, there is no exact analog to this for the rich set of unsupervised analysis techniques that do not fit a likelihood function to the data. 

Extrapolating from the generative case, it would be valuable to associate data and a family of models with a typical range of values that can be considered to be sufficiently close to the original model. One needs a sense of the empirical stability of the model - the measure of the degree to which the original modeling procedures can generate different results under resampling. If this property is understood, then it is possible to treat the question of retraining as an outlier detection problem over the distance distribution of the unsupervised model. This can be thought of as analogous to the shape distribution approach proposed by Osada \cite{osada2002shape} for object recognition. However, to do this requires a measure of model distance or similarity. 

One family of unsupervised techniques for data analysis that has generated excitement in recent years is Topological Data Analysis \cite{carlsson2009topology}. Data sets can vary widely in complexity and inherent structure. Genomic data will naturally have different structure compared to time series or text data. Advances in topological data analysis have presented researchers with techniques for analyzing point cloud data and uncovering that inherent structure. One such approach is the Mapper algorithm introduced by Singh et al \cite{singh2007topological}. Mapper has seen success in a wide variety of applications. In the biological sciences and medicine, it has been used for precision oncology \cite{prec_onc}, for developing new insights into the structure of the brain \cite{brainmap}, for discovering new sub-types of diabetes \cite{diabetes}, for analyzing patterns in asthma patients \cite{asthma}, and more. It has also found usage in remote sensing \cite{remotesensing}, analysis of geological patterns \cite{earthquakes}, and natural language processing  \cite{doshi2018movie}.  

However, as with other unsupervised approaches, evaluating the empirical stability of Mapper graphs is non-trivial. Carri\'ere, et al \cite{carriere2017structure} have previously analyzed the stability of the 1-d Mapper algorithm with respect to convergence to the Reeb graph while focusing largely on the topological features (loops and flares); however, the question of the extent to which an individual Mapper model trained on sample data will vary under resampling with respect to the metric structure of the point cloud remains unaddressed. 
  
The first step in evaluating the empirical stability is to develop a distance measure between Mapper graphs. Existing approaches for related structures like graphs or simplicial complexes do not adequately capture variation between Mapper graphs because they largely focus either on the extrinsic distances in the embedding metric space or intrinsic distances captured by the graph structure and fail to capture both aspects concurrently. 

\textbf{Contributions} \label{contribs}In this work, we develop a new framework for comparison of Mapper graphs by analyzing them as graph structured metric-measure spaces embedded in extrinsic metric spaces. We develop an optimal transport based metric\footnote{Code to be available at \url{https://github.com/mikemccabe210/mapper_comparison}} for evaluating the distance between Mapper graphs. Furthermore, we utilize this metric to describe the empirical stability of Mapper graphs and use this information to indicate changes in the population used to generate Mapper graphs. While this paper focuses specifically on Mapper graphs and other nerve complex representations, we also demonstrate that the technique is sufficiently general to apply to arbitrary weighted undirected graphs. 

\textbf{Related Work} Graph comparison is a field with a long history for which several excellent surveys are available \cite{akoglu2015graph, ranshous2015anomaly}. One important distinction in the graph comparison literature exists between techniques for graphs with known node correspondences and techniques for those without. We focus on techniques that do not require known node correspondences as nodes in Mapper graphs correspond to open sets of points and therefore node correspondences only exist in the case that multiple graphs are built from the same data. Techniques for analyzing graphs without known correspondences include spectral distances \cite{bunke2007spectral,shoubridge2002spectral} and NetSimile \cite{netsimile}. For attributed graphs, Wasserstein metrics can also be used.

While optimal transport over graph structured data is a well studied problem, we focus on techniques utilizing optimal transport specifically for comparing two or more sets of graph structured data.  M\'emoli \cite{memoli2011gromov} introduced the Gromov-Wasserstein distance as a Wasserstein variant for object recognition.  Hendrikson \cite{hendrikson2016gromov} continued that line of inquiry and demonstrated the use of Gromov-Wasserstein for undirected graphs. Chowdury and M\'emoli \cite{Chowdhury2018gromov} introduced an adaptation of the Gromov-Wasserstein metric for comparing directed graphs. Peyr\'e, Cuturi, and Solomon \cite{peyre2016gromov} demonstrated an efficient approximate Gromov-Wasserstein discrepancy using Sinkhorn iterations. Vayer, et al \cite{vayer2018fused} identified the need for metrics capturing both intrinsic and extrinsic distances and introduced the fused Gromov-Wasserstein distance. Following a different path to address to problem of transport between sets with intrinsic structure, Alvarez-Melis \cite{melis17submodtransport} used a submodular loss function to promote transport between similar topological features.  

Additionally, there have been several previously explored approaches to measure the stability of the Mapper algorithm. Carri\'ere and Oudot \cite{carriere2017structure} demonstrated that Mapper converges to the Reeb graph and furthermore demonstrated that it is possible to use this convergence property to tune Mapper parameters \cite{oudottuning} using topological persistence. Dey, M\'emoli, and Wang \cite{dey2016multiscale} introduced the Multi-scale Mapper, which aims to improve the stability of the Mapper algorithm by using a tower of covers rather than a single cover in order to capture information at multiple scales.

\section{Background}
\subsection{Mapper}\label{mapper}

Let $(X, d)$ be a metric space over the set $X$ endowed with metric $d$ hereafter referred to as $X$. A cover of $X$ is a family of subsets $U = \{U_{\alpha}\}_{\alpha \in A}$ where $A$ is the indexing set of $U$ such that $X = \cup_{\alpha \in A} U_\alpha$. The nerve of the covering is the simplicial complex $N(U)$ whose vertex set is the indexing set $A$ and where the family ${\alpha_0, \alpha_1, \dots, \alpha_k}$ spans a $k$-simplex in $N(U)$ if and only if $U_{\alpha_0} \cap U_{\alpha_1} \cap \dots U_{\alpha_k} \ne \emptyset $. 
\begin{figure}[htp] 
	\centering
	\includegraphics[width=.24\textwidth]{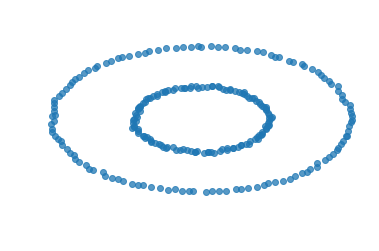}
	\includegraphics[width=.24\textwidth]{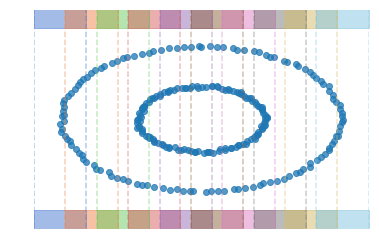}
	\includegraphics[width=.24\textwidth]{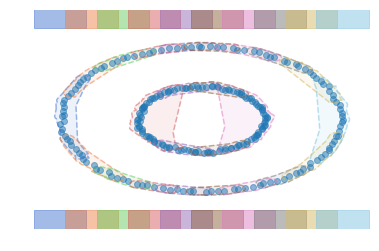}
	\includegraphics[width=.24\textwidth]{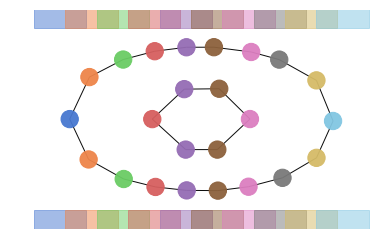}
	
	\caption{From left to right: (a) Concentric circles embedded in $\mathbb{R}^2$. (b) Cover defined as uniform spaces over the function $f : X \times Y \to X$. (c) Pullback cover refined by clustering within each open subset. (d) Output is nerve of the refined cover, which captures the topology of the original data in a condensed form.}
	\label{figmapper}
\end{figure}   

Mapper is an algorithm for deriving a visual topological summary of a dataset. Given a data set $X$ and a continuous function $f : x\to \mathbb{R}$, the Mapper algorithm first creates an open cover over the image of $f$. This cover is then pulled back into $X$ and refined by performing partial clustering within the open sets of the pullback cover. The output of the algorithm is a simplicial complex formed by the nerve of the refined pullback cover, which can be thought of as a discrete analog to a Reeb graph  \cite{carriere2017structure}. This process can be seen in Figure \ref{figmapper}. In (d), the Mapper graph is laid out in the ambient metric space according to the original metric, though Mapper is generally used to visualize metric spaces that can not be so easily embedded in $\mathbb{R}^2$.

The key parameters for Mapper are the metric $d$, the filter functions (often called the lenses) and the procedure for generating a cover from those lenses. One typical procedure is to generate a set of uniformly spaced overlapping bins over the range of each lens function in $\mathbb{R}$ as the initial cover. The number of bins over the given lens is referred to as the \textit{resolution} and the degree of overlap is called the \textit{gain}. 

The Mapper graph $M(U, \mathcal{G}, d)$) is represented as an open cover $U$ of $X$, the graph $\mathcal{G}$ formed by the nerve of that cover, and the metric $d$ defined over the metric space $X$ . Each vertex corresponds to one open set in the cover such that we can define a map $\phi : V \to U$ where $\phi(v_\alpha) = \{x_i\ |\ x_i \in U_\alpha\}$.

\subsection{Optimal Transport}\label{ot}
Optimal Transport distances provide an elegant way of comparing distributions by associating transport between different bins of a histogram with a cost function. Also called Wasserstein distances or Earth Mover’s Distances, metrics based on optimal transport theory have found a wide variety of applications in machine learning. While they have been long used by the Computer Vision community \cite{rubner1998metric}, their use has recently expanded to other domains like natural language processing \cite{kusner2015word}, domain adaptation \cite{courty2017optimal}, and deep learning  \cite{arjovsky2017wasserstein}. 

\subsubsection{Wasserstein Distance}
Wasserstein Distances can be thought of as distances over subsets defined on metric-measure spaces (mm-spaces). A metric-measure space is a triple $(X, d_X, \mu_X)$ where $(X, d_X)$ is a metric space and $\mu_X$ is a Borel measure with support $X$. 

Assume that $X$ and $Y$ are compact mm-spaces and let $\mu_X$ and $\mu_Y$ be Borel measures defined with supports $X$, $Y$ respectively. Given a cost function $c$, the Kantorovich formulation of optimal transport seeks to identify a transportation plan $\mu : X \to Y$ that solves:
\begin{equation*}
\inf_\mu \bigg\{ (\int_{X \times Y} c(x, y)\ d\mu(x, y)^p\ )^{1/p}\  \bigg |\ \mu \in \mathcal{M}(\mu_X, \mu_Y)\bigg\}
\end{equation*}
where $\mathcal{M}(\mu_X, \mu_Y)$ is the set of valid transportation plans, which are defined as
 joint distributions with marginals $\mu_X$ and $\mu_Y$. Wasserstein distances in general are parameterized by $p$, but for simplicity of notation, the remainder of the paper will assume that $p=1$. In the discrete case, we can represent $\mathcal{M}$ as the polyhedron:
 \begin{align*}
 \mathcal{M}(\mu_X, \mu_Y) := \{\mu \in \mathbb{R}^{n \times m}_+\ |\mu \textbf{1} = \mu_Y,\ \mu^T \textbf{1} = \mu_X\}
 \end{align*}
 and $C$ as the matrix of pairwise cost values, which leads to the discrete transport problem:
\begin{align*}
&\min_\mu \sum_{(i,j)\in X \times Y} \mu_{ij}C_{ij} \\ 
& s.t. \ \mu \in \mathcal{M}(\mu_X, \mu_Y)
\end{align*}
This linear program can be solved exactly in $O(n^3\log(n))$ time using interior point methods \cite{orlin1993faster}. Alternatively, one can calculate fast, approximate Wasserstein distances using Sinkhorn iterations \cite{cuturi2013sinkhorn}. 

The Wasserstein distance has proven to be an effective method for measuring distances between subsets of a mm-space; however, the topological features of two different graphs cannot be directly compared over the mm-space because distances over graphs are only defined intrinsically, meaning that two graphs are entirely different mm-spaces even if they are defined over the same extrinsic metric space. Comparing graphs using standard Wasserstein distances over intrinsic distances would require defining edges between the two graphs, in which case it may not be feasible to demonstrate that the resulting discrepancy is still a metric. 

\subsubsection{Gromov-Wasserstein Distance}
In 2011, M\'emoli introduced the Gromov-Wasserstein distance for coordinate independent matching \cite{memoli2011gromov}. Rather than measuring extrinsic distances within a mm-space the Gromov-Wasserstein distance is defined \textit{between} mm-spaces. It measures how far two mm-spaces are from isomorphism. Two mm-spaces $X,\ Y$ are isometric if there exists a surjective function $\psi: X \to Y$ such that $d_X(x,x')=d_Y(\psi (x), \psi (x'))$. The spaces are isomorphic if it is also true that $\psi_\# \mu_X = \mu_Y$. 

Assume we have two compact mm-spaces $(X, d_X, \mu_X)$, $(Y, d_Y, \mu_Y)$. Let $\Gamma_{X, Y} \in X\times X \times Y \times Y $ be an order four tensor defined as:
\begin{equation*}
\Gamma_{X, Y}(x, x', y, y') = |d_X(x, x') - d_Y(y, y') |
\end{equation*}

$\Gamma_{X, Y}$ contains the pairwise differences in intrinsic distances between pairs of points $X$ and $Y$.  The discrete Gromov-Wasserstein metric is then defined as:

\begin{equation*}
\min_\mu \bigg\{\sum_{x_i, x_{i'} \in X}\sum_{y_j,y_{j'} \in Y} \Gamma_{X, Y}(x_i, x_{i'}, y_j, y_{j'})\mu_{ij}\mu_{i'j'}\ \bigg  | \ \mu \in \mathcal{M}(\mu_X, \mu_Y) \bigg\}
\end{equation*}

The Gromov-Wasserstein metric can be shown to converge to a local minima using the alternating linear program described by M\'emoli \cite{memoli2011gromov} and documented in Henrikson \cite{hendrikson2016gromov}.

The Gromov-Wasserstein distance has been used for shape recognition and arbitrary graph comparison \cite{Chowdhury2018gromov, hendrikson2016gromov}. For these use cases, coordinate independence is a benefit of the Gromov-Wasserstein metric. There are known invariances under translation, rotation, and often under certain types of deformation. While Mapper graphs inherit many of those properties in certain contexts, it cannot be assumed that these properties hold for all potential applications. 

As an example, consider the yearly trends for two businesses. One would expect to see clear cyclic patterns, which would manifest topologically as a pair of loops. However, the absolute coordinates still give us valuable information when it comes to comparing these cycles. If one business is significantly larger, we would expect for these two cycles to be quite far within a reasonable metric space that accounts for financial value.

Vayer et al \cite{vayer2018fused} demonstrated a hybrid metric titled the Fused Gromov-Wasserstein distance that jointly solves the optimal transport problem for both the Gromov-Wasserstein and traditional Wasserstein costs. We define the elements in $C \in R_+^{n \times m}$ to be:
\begin{equation*}
C_{ij} = d_E(x_i, y_j)\quad \forall\ x_i\in X,\ y_j\in Y 
\end{equation*}

where $d_E$ is the extrinsic metric defined on our embedding space. We can then define the fused Gromov-Wasserstein distance as:

\begin{equation*}
\min_\mu \bigg\{\sum_{x_i, x_{i'} \in X}\sum_{y_j,y_{j'} \in Y} [\Gamma_{X, Y}(x_i, x_{i'}, y_j, y_{j'}) + C_{ij}]\mu_{ij}\mu_{i'j'}\ \bigg  | \ \mu \in \mathcal{M}(\mu_X, \mu_Y) \bigg\}
\end{equation*}

Vayer demonstrated that the fused Gromov-Wasserstein distance can be computed using Conditional Gradient methods or through an entropic-regularized Sinkhorn iteration scheme.

However, this metric requires each intrinsic metric space to be bounded which is not the case by convention for graphs which are not path-connected. It also shares the computational cost of the Gromov-Wasserstein distance, which is a non-convex quadratic problem. If we plan to use the metric in an outlier detection setting, we will need to evaluate the distance over a large number of networks in order to establish a baseline. One way to ease the computation burden might be to use Sinkhorn iterations, but the entropic-regularized Gromov-Wasserstein metric has been observed to sometimes require over-regularization in order to converge \cite{vayer2018fused}. These over-regularized solutions can fail to capture interesting metric structure. 

\section{Network Augmented Wasserstein Distance }
\subsection{Intuition}
Before measuring the expected variation in Mapper graphs, we must first define a method to measure distance between Mapper graphs. Following in the footsteps of Koutra, et al. \cite{koutra2013deltacon}, we enumerate a set of properties that we would like an intuitive metric to maintain when comparing graphs. Here, we adapt Koutra's graph comparison criteria from the case of general graph comparison to the case of comparing Mapper graphs specifically: 
\begin{enumerate}\label{crits}
	\item \textit{Edge Importance} - Changes that create disconnected components in high density regions should be penalized more than those that disconnect low density regions or those than do not alter the path-connectedness of the graph.
	\item \textit{Metric Awareness} - Operations that preserve structure, but translate the graph within the extrinsic space should be proportional in distance to the size of the translation. Furthermore, the cost of adding or removing edges should be related to the extrinsic distance spanned by the edge. 
	\item \textit{Multi-scale} - Networks with different resolutions, but similar topological structure should be considered "close". Collapsing or expanding individual vertices in a manner that does not fundamentally alter topological structure or neighborhood information in the underlying metric space should be smaller changes relative to modifications that do alter said structure. 
	\item \textit{Density Awareness} -  Density changes without changes in graph structure are still significant and should be reflected in the metric. 
\end{enumerate}
\indent There are a two natural approaches here. Given that the output of the Mapper algorithm is a graph reflecting the underlying topology of the set, it makes sense to examine standard graph distance measures that do not require known node correspondences. These include spectral distances and node featurization approaches. However, the challenge with approaching the Mapper graph comparison problem from a purely graph theoretic or graph mining approach is that the graphs analyzed by these techniques generally do not capture information about the underlying metric space. 

Alternatively, given that our graph can also be interpreted as a simplicial complex defined to capture the topology of the space, it might make sense to analyze mapper graph similarity using topological persistence \cite{zomorodian2005computing, edelsbrunner2008persistent}. Persistence captures the way topological features change over different scales. Using persistence, we could capture some aspects of the metric structure, but we would capture very little information about sample densities or the ambient metric space.

\paragraph{Mapper graphs as metric-measure spaces}The approach that we take in this paper is representing Mapper graphs as graph structured metric-measure spaces (mm-space) that are also imbued with extrinsic metrics. To define our mm-space, we'll need a measure $\mu$ and a metric $d$ defined over the mapper graph $M$. Each Mapper graph is parameterized by a metric space and can represent density information about the sample over that metric space. 

Recall that $\phi$ is a function that maps each vertex $v_i \in V$ to an open set $U_\alpha \in U$ which contains one or more  $x_i \in (X, d)$. We will use this fact to define our measure $\mu_X$ as:
\begin{equation*}
\mu(v_i) = \frac{|\phi(v_i)|}{\sum_{v_i \in V} |\phi(v_i)|}
\end{equation*}
While the choice of distance between vertices is flexible, one common approach for measuring a distance between sets is the Hausdorff distance \cite{RockWets98}. If $d$ is a metric, then the Hausdorff distance:
\begin{equation*}
d_H = \max \bigg(\sup_{x \in X} \inf_{y \in Y} d(x, y), \sup_{y \in Y} \inf_{x \in X} d(x, y)\bigg)
\end{equation*}
is also a metric. 

Given the metric defined between nodes and their corresponding open sets and the measure defined over the Mapper graph, we can use tools from optimal transport to define a metric between our mm-spaces. 

\subsection{Network Augmented Wasserstein Distance}
While the fused Gromov-Wasserstein metric captures the phenomenon that we’re interested in, it breaks apart in the case of graphs with multiple connected components.  For that, we need to introduce an approximation that is applicable to the more general case which we call the Network Augmented Wasserstein (NAW) Distance. 

In the 2011 paper, M\'emoli demonstrated a series of lower bounds for the Gromov-Wasserstein metric. The first of these lower bounds, aptly called the First Lower Bound (FLB), is derived using the eccentricity of the vertices, which M\'emoli defines as $s_{X}(x) = \sum_{x_i \in X}\mu(x_i)d_X(x, x_i)$. Using $\mathcal{C}(x_i)$ to denote the set vertices in the same connected component as $x_i$, we adopt a commonly used non-standard definition of eccentricity that does not go to infinity in the case of disconnected components, $s_{X}(x) = \sum_{x_i \in \mathcal{C}(x_i)}\mu(x_i)d_X(x, x_i)$. In the case of a fully path-connected graph, the two definitions are equivalent.

Given a pair of metric-measure spaces, $X$, $Y$, M\'emoli's first lower bound (FLB) is computed by solving the transport problem:
\begin{equation*}
\min_{\mu \in \mathcal{M}(\mu_X, \mu_Y)} \sum_{x_i\in X, y_j\in Y} |s_X(x_i) - s_Y(y_j)|\mu_{ij}
\end{equation*}

Transportation distances are known to be metrics conditional upon the measures being probability density functions and the cost function being a metric. The difference in eccentricities is equivalent to a Minkowski distance between feature spaces in $\mathbb{R}$ defined by the eccentricity of the nodes, making the FLB a metric. 

To adapt the FLB to the fused Gromov-Wasserstein case, we need to reintroduce the ambient metric space into which both of our metric-measure spaces are embedded:

\begin{equation*}
\min_{\mu \in \mathcal{M}(\mu_X, \mu_Y)} \frac{1}{2}\sum_{x_i\in X, y_j\in Y} (|s_X(x_i) - s_Y(y_j)| + d_E(x_i, x_j)) \mu_{ij}
\end{equation*}

which can be demonstrated to be a metric by the observation that it is an affine combination of a Minkowski distance over a feature space in $\mathbb{R}$ defined by the eccentricities and an arbitrary secondary metric. This new metric is a straightforward transportation problem that can be solved either as a linear program or approximated using Sinkhorn iterations. 

In the case of graphs with a single connected component, this formulation constitutes a lower bound on the fused Gromov-Wasserstein distance in the same manner that the original FLB constituted a lower bound on the Gromov-Wasserstein distance. In the case of pure singleton clusters with no additional topological information, it reduces to a standard Wasserstein distance over the ambient metric space.

\section{Experiments}
We aim to demonstrate that distances between Mapper graphs using the proposed metric fulfill the criteria enumerated in section \ref{crits} for an intuitive distance over Mapper graphs. First, we evaluate our proposed metric against a set of alternative distance and similarity measures on synthetic data to demonstrate that the proposed method better satisfies the desired properties of a Mapper graph similarity compared to existing methods.  We then evaluate the efficacy of the proposed metric for several applications both using Mapper graphs and using other graph-structured data.
\subsection{Metric Validation}
For each case, we build a reference graph on a synthetic data set, and three follow-up Mapper graphs that have been altered in accordance with the given axiom such that there is a natural ordinal ranking of the networks. We choose these synthetic data sets for their visually obvious structure in the Euclidean plane. For constructing the Mapper graphs, the Euclidean metric will be used and the lens function will be the projection onto the $x$ axis. For these tests, the following graph distances will be used: 
\begin{itemize}
	\item NAW Distance
	\item Wasserstein Distance
	\item $\lambda$-Distance (Laplacian, Adjacency) \cite{shoubridge2002spectral}
	\item NetSimile \cite{netsimile}
\end{itemize}

The Python libraries netcomp and Python Optimal Transport (POT) \cite{flamary2017pot} were used for graph comparison algorithms and optimal transport solutions respectively. 

One important note is that to impose edits on Mapper graphs, we must add a member of the cover corresponding to one vertex to the cover corresponding to another vertex. This is not purely a structural change as it also impacts the Hausdorff distance measured between the sets corresponding to the vertices. 

\subsubsection{Edge Importance} 
\begin{figure}[htp] 
	\centering
	\includegraphics[width=.24\textwidth]{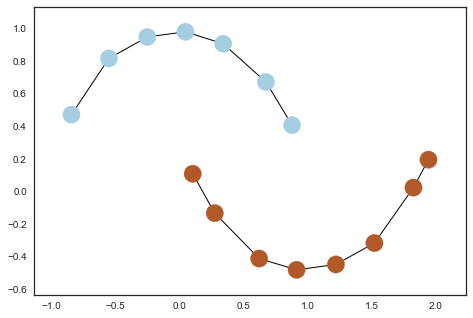}
	\includegraphics[width=.24\textwidth]{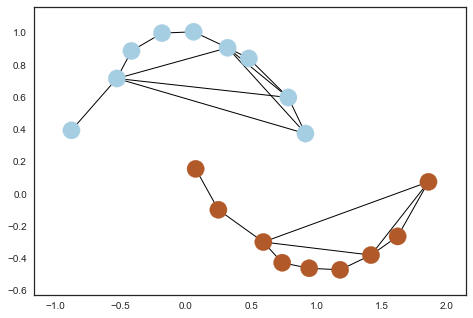}
	\includegraphics[width=.24\textwidth]{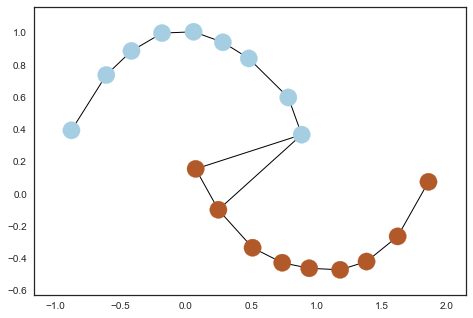}
	\includegraphics[width=.24\textwidth]{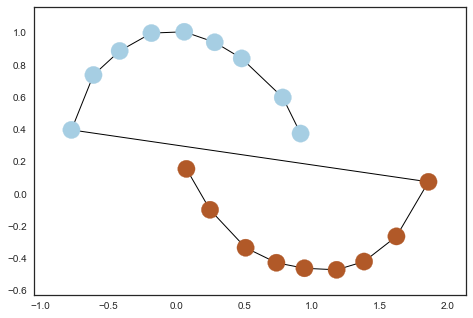}
	
	\caption{Edge Aware- From left to right: (a) Mapper graph of crescent moon data (b) Graph altered by adding multiple edges to each connected component (c) Single edge added connecting components between the near end points (d) Single edge added between far endpoints}
	\label{edgeaware}
\end{figure}

To evaluate edge importance, we use the twin moon data set. In \autoref{edgeaware}, we demonstrate a series of alterations to the graph structure with minimal density or metric changes over the graph. In the first, we add edges between nearby nodes within the same components. In the second, we connect the components by adding an edge between nodes consisting of points that are close within the metric space. Finally, we connect the components by adding an edge that crosses the space. 

A distance measure that matches our intuition would penalize the connections between components much more heavily than the connections within existing components. Furthermore, since Mapper graphs are embedded in Euclidean space, we would expect edges that span large distances in the Euclidean plane to lead to correspondingly larger distances.

\subsubsection{Metric Awareness}  
\begin{figure}[htp] 
	\centering
	\includegraphics[width=.24\textwidth]{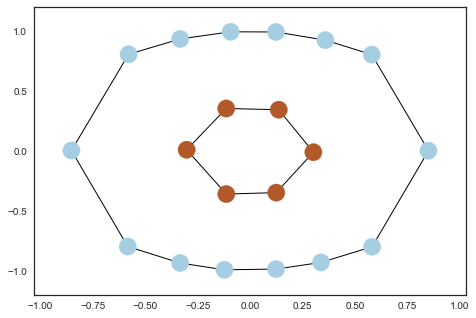}
	\includegraphics[width=.24\textwidth]{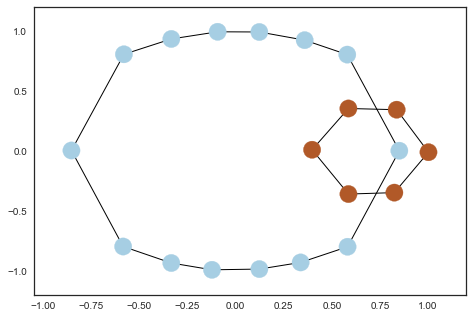}
	\includegraphics[width=.24\textwidth]{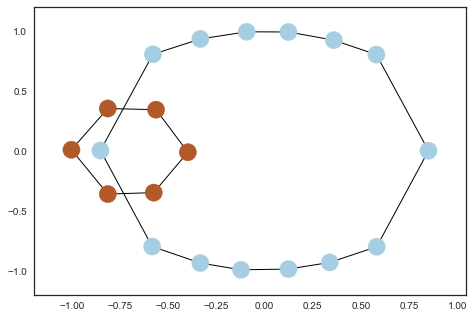}
	\includegraphics[width=.24\textwidth]{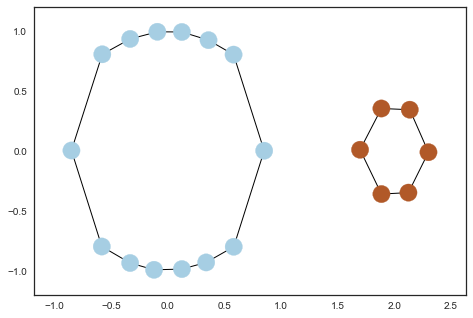}
	
	\caption{Metric Aware - From left to right: (a) Mapper graph of concentric circles dataset (b) Interior circle translated right (c) Interior circle translated left (d) Interior circle translated outside of graph}
	\label{metricaware}
\end{figure}

We use concentric circle data in Euclidean space to measure metric awareness. In each successive frame in \autoref{metricaware}, we apply translations of different size and direction to our inner circle. No changes are made to the graph structure or intrinsic metric structure. Graphs (b) and (c) are translated by the same value in opposite directions while in (d) the inner circle has been translated completely outside of the larger circle.

Intuitively, (b) and (c) would be expected to have approximately the same distance/similarity values with the original graph while (d) would be expected to be significantly further. As we do not actually make changes to the connectivity structure of the graph, one can assume that the pure graph distances will not register the changes made in this experiment.

\subsubsection{Multi-scale} We return to the twin moons data to evaluate the scale sensitivity of the metrics. We alter our initial network by halving the resolution, doubling the resolution, and finally dropping resolution to one so that the graphs collapsed to a single point. 
\begin{figure}[htp] 
	\centering
	\includegraphics[width=.24\textwidth]{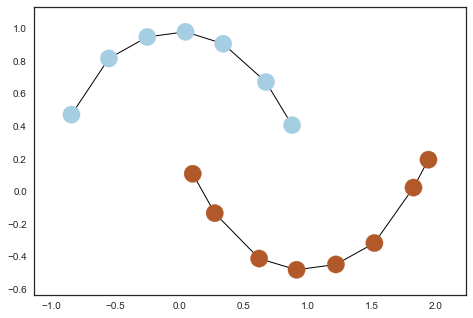}
	\includegraphics[width=.24\textwidth]{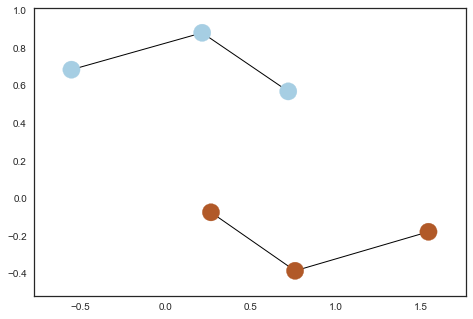}
	\includegraphics[width=.24\textwidth]{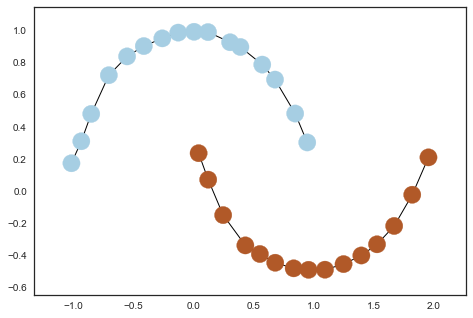}
	\includegraphics[width=.24\textwidth]{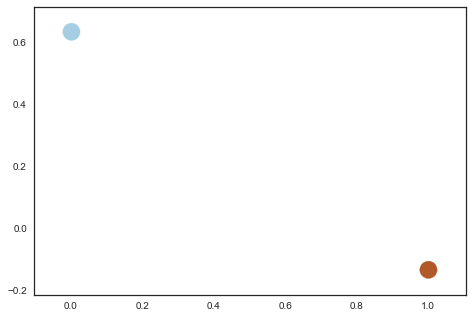}
	
	\caption{Multi-scale - From left to right: (a) Mapper graph of twin moons dataset (b) Mapper graph at half resolution (c) Mapper graph at double resolution (d) Mapper graph collapsed into two clusters}
	\label{multiscale}
\end{figure}   

In this test, we'd expect significant distances on the pure graph measures. In fact, they might be expected to overpenalize the changes, as without metric and density information, it is difficult to determine that the Mapper graphs are representing the same underlying metric space.

\subsubsection{Density Awareness} In this experiment, we maintain a single network structure, but arbitrarily move a fixed density from the outer ring to different locations in the graph without altering the intrinsic or extrinsic metric spaces. Graph (a) is built from concentric circles where the density is split evenly between the two circles and uniformly distributed within each circle. In (b), we localize the density of the outer circle so that we have a high density region at the top of the circle. In (c), we move that density into the interior ring. In (d) we introduce a new connected component located at the mean coordinate value with density equivalent to the density removed from the outer ring.

\begin{figure}[htp] 
	\centering
	\includegraphics[width=.24\textwidth]{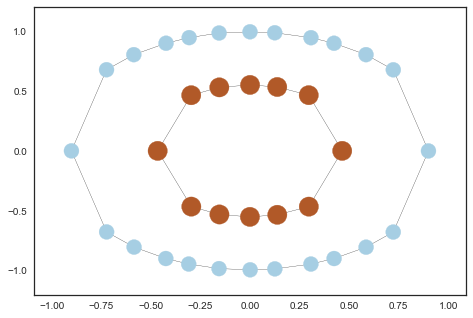}
	\includegraphics[width=.24\textwidth]{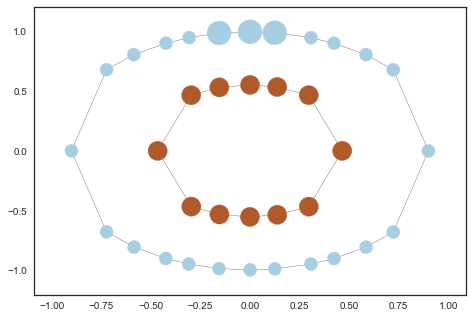}
	\includegraphics[width=.24\textwidth]{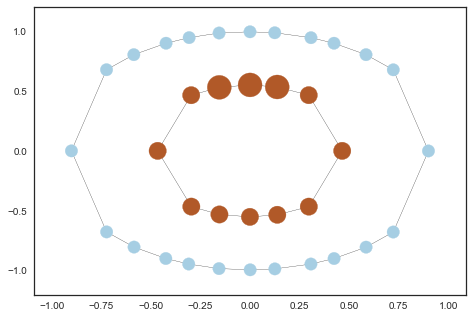}
	\includegraphics[width=.24\textwidth]{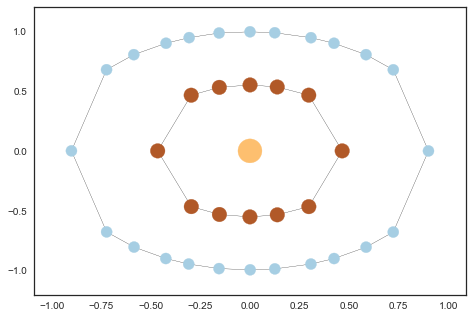}
	
	\caption{Density Aware - From left to right: (a) Mapper graph of concentric circles data with probability mass split evenly over the circles and uniformly distributed over each circle (b) Density redistributed from outer circle to be concentrated at a point on the outer circle (c) Density redistributed from the outer circle to be concentrated at a point on the inner circle (d) Density redistributed to be concentrated at a new point at the mean coordinate value}
	\label{densityaware}
\end{figure}   

We would expect that reallocating mass along the outer ring would be less significant than relocating it to a new connected component. Despite being centrally located in the ambient space, we would expect that the significantly different intrinsic structure of our new density region in (d) should result in a larger change. 

\subsubsection{Metric Validation Results}

\begin{table}[h!]
	\centering
	\begin{tabular}{||c ||c c c c c ||} 
		\hline
		\multicolumn{1}{||c||}{} & \multicolumn{5}{c||}{Edge Awareness} \\[0.5ex] 
		\hline
		Networks & NAW & Wasserstein & $\lambda$-Lap & $\lambda$-Adj & NetSimile   \\ [0.5ex] 
		\hline\hline
		(1, 2) &0.236 &0.226  & 4.813 & 2.130& 18.537   \\
		(1, 3) & 0.988 & 0.055 & 1.221 &0.754 & 20.156  \\
		(1, 4) &  1.994  & 0.220 & 0.739 &  0.534 & 15.751 \\ [1ex] 
		\hline
		\hline
		\multicolumn{1}{||c||}{} & \multicolumn{5}{c||}{Metric Awareness} \\[0.5ex]
		\hline\hline
		(1, 2) & 0.177 & 0.355 & 0 & 0 & 0  \\
		(1, 3) & 0.177 & 0.355 &0 &0 & 0  \\
		(1, 4) & 0.507 & 1.015 & 0 & 0 & 0  \\ [1ex] 
		\hline
		\hline
		\multicolumn{1}{||c||}{} & \multicolumn{5}{c||}{Multi-scale} \\[0.5ex]

		\hline\hline
		(1, 2) & 0.417 & 0.643 & 5.897 & 4.535 & 15.676  \\
		(1, 3) & 0.187 &0.327 & 7.287 & 6.305& 16.682  \\
		(1, 4) & 0.994 & 1.418 & 8.602 & 5.616 & 26.615  \\ [1ex] 
		\hline
		\hline
		\multicolumn{1}{||c||}{} & \multicolumn{5}{c||}{Density Awareness} \\[0.5ex] 

		\hline\hline
		(1, 2) & 0.172 &0.200& 0 & 0 & 0  \\
		(1, 3) & 0.191 & 0.153 & 0 &0 & 0  \\
		(1, 4) & 0.211 & 0.150 & \textasciitilde 0 & 1.348 & 15.385  \\ [1ex] 
		\hline

	\end{tabular}
	\caption{Unnormalized metric validation results}
	\label{results}
\end{table}
In \autoref{results}, we see that the results largely match our expectations. The $\lambda$-distances do very well in discriminating between graph changes of different magnitudes as can be seen in the edge awareness case. However, approaches that only examine graph structure predictably fail to capture variation when graph structure goes unchanged. More general approaches such as the NAW and Wasserstein distances are not hampered in these situations as they are evaluating the underlying metric space and not the graph structure.

On the other hand, the Wasserstein distance struggles on the edge awareness case due to the fact that there is very little change in the metric structure - we are only moving a small number of points between sets to create new edges. It also struggles in the density awareness experiments due to the fact that it fails to account for cases where density is being moved to an entirely different component. 

Only the NAW distance performs adequately on all stated criteria. However, even the NAW struggles on the Multi-scale task as it indicates an expansion is a significantly larger change than a contraction, while intuitively they should be similar changes. 
\subsection{Applications} In the validation section, we demonstrated that our metric satisfies certain intuitive criteria that make it suitable for comparing Mapper graphs. In this section, we apply the proposed metric to several real datasets to demonstrate that the metric is useful in practice as well. 

\subsubsection{Model Drift and Mapper} 

The motivating problem for this research was the challenge of determining when a Mapper model no longer fits the data. In this section, we evaluate the ability to identify to introduction of a new class by using the distance distribution over the NAW distance. Using Fashion MNIST \cite{fmnist}, we begin by taking a sample of size $n$ with replacement from the full data set while excluding a single class. We then take one hundred additional samples with replacement of size $n$ from the same subset of the data set. These samples are used to determine the medioid Mapper graph and the distribution of distances from that medioid graph. We then take ten additional samples of the same size from a subset of the data set containing all ten classes. 
\begin{figure}[htp] 
	\centering
	\includegraphics[width=.49\textwidth]{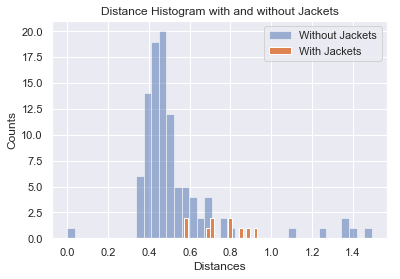}
	\includegraphics[width=.49\textwidth]{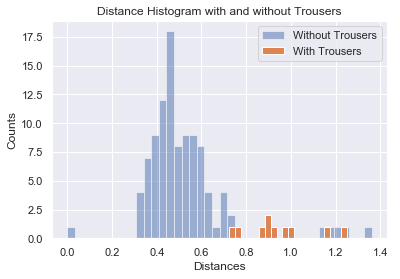}
	\caption{Distance distributions for the jacket and trouser experiments.}
	\label{hists}
\end{figure}

This is first performed on the trouser class. As the majority of classes in Fashion MNIST are upper body garments, trousers are naturally an outlier in this set. We then evaluate performance on jackets, which have much higher similarity to the other upper body garments in the data set. The distance distributions for both experiments can be seen in \autoref{hists}.

\begin{figure}[htp] 
	\centering
	\includegraphics[width=.32\textwidth]{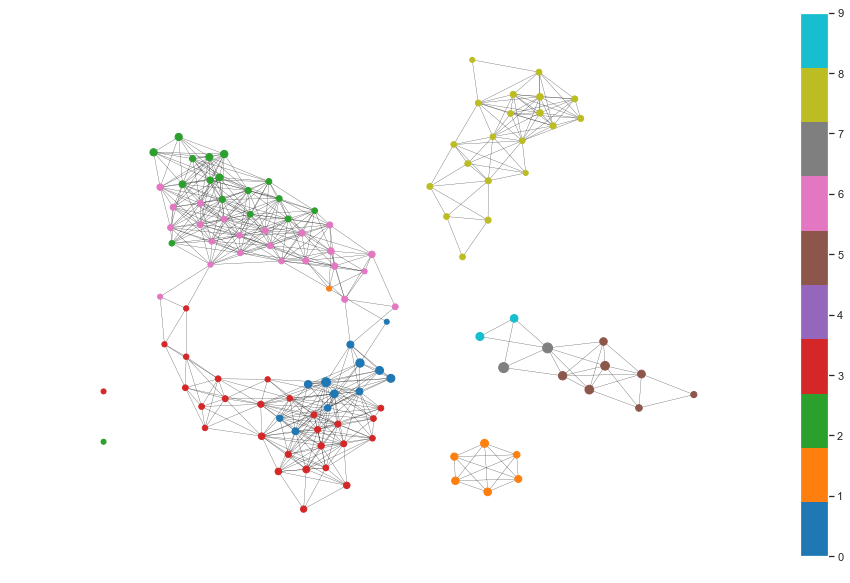}
	\includegraphics[width=.32\textwidth]{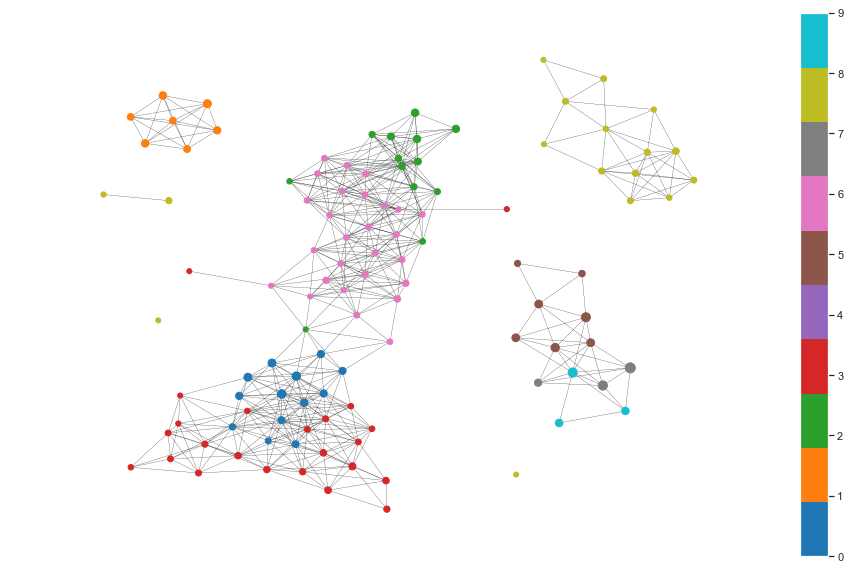}
	\includegraphics[width=.32\textwidth]{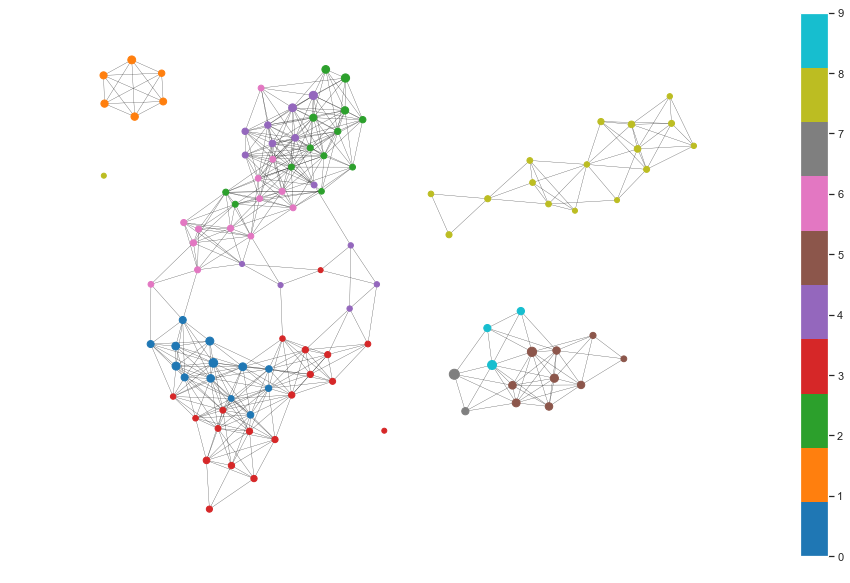}
	\caption{With and without jackets - From left to right: (a) Medioid graph from base distribution (b) Base graph with highest distance from medioid. (c) Anomalous sample closest to medioid }
	\label{jacketdists}
\end{figure}

While in both scenarios our samples with the "anomaly" class are clearly on the higher end of the distance distributions, it might be surprising that several of our samples containing jackets are not clear outliers. However, when we look at the actual graphs (\autoref{jacketdists}), the reason becomes clear.

The Mapper structure of our data set appears to contain a relatively unstable loop in the largest component featuring our upper body garments. In the far left graph, we can see that our medioid has a tenuous connection completing the graph. In the center, we can see the sample without the jacket class that is the furthest from our medioid. In this example, the connection is broken, resulting in the loss of the prevailing loop component, which is reflected in the metric. On the right, the addition of the jacket class appears to have made our loop sturdier and reinforced it with additional connections. So while our samples with jackets are noticeably further from the medioid than the majority of our non-anomalous samples, our loops are less likely to break, resulting in fewer extreme outliers.
 \begin{figure}[htp] 
	\centering
	\begin{subfigure}[t]{.32\textwidth}
		\centering
		\includegraphics[width=\textwidth]{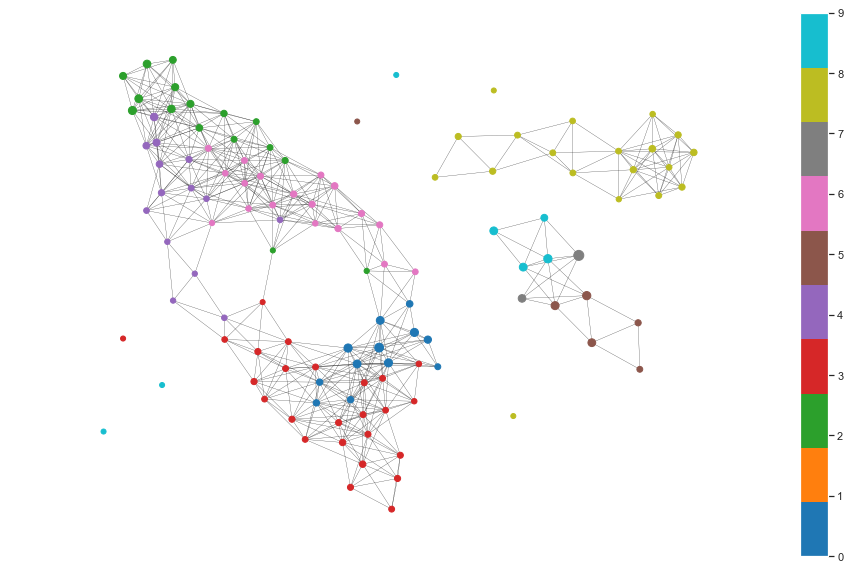}
		
	\end{subfigure}
	\begin{subfigure}[t]{.32\textwidth}
		\centering
		\includegraphics[width=\textwidth]{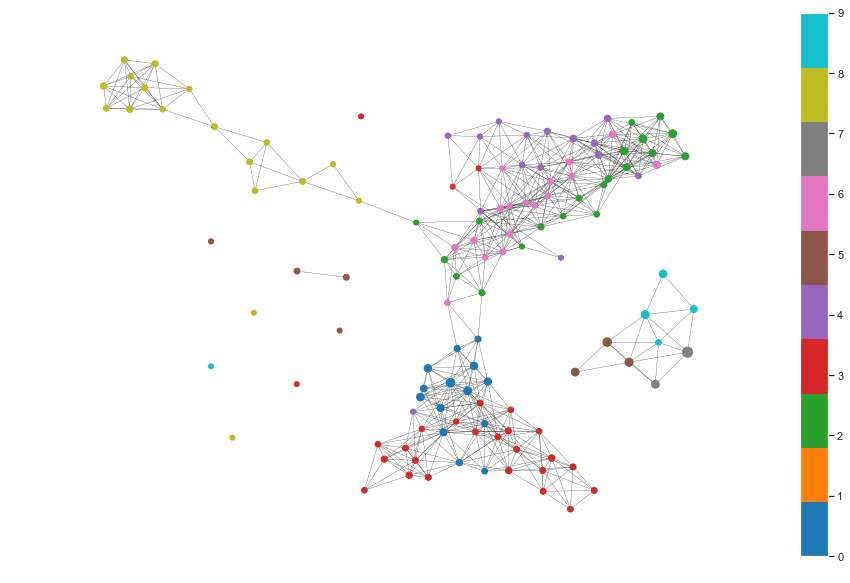}
		
	\end{subfigure}
	\begin{subfigure}[t]{.32\textwidth}
		\centering
		\includegraphics[width=\textwidth]{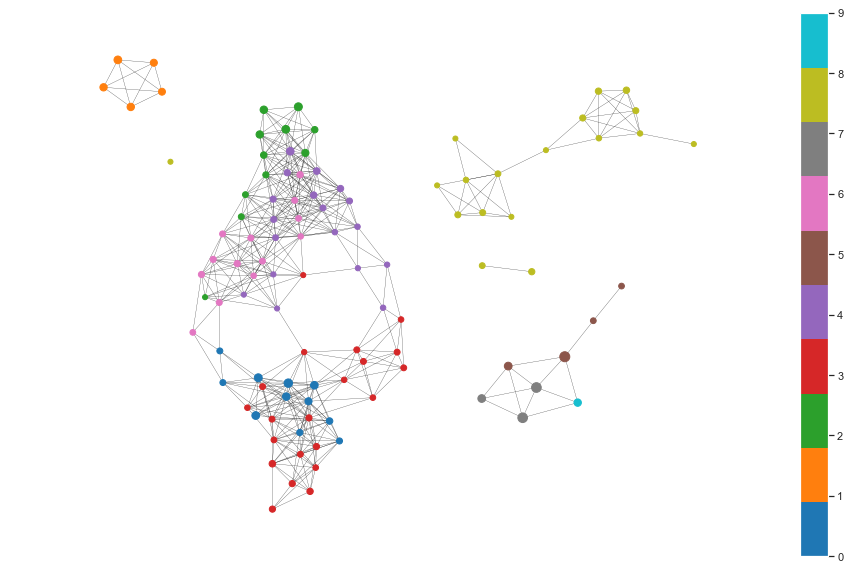}
		
	\end{subfigure}
	\caption{With and without trousers - From left to right: (a) Medioid graph from base distribution (b) Base graph with highest distance from medioid. (c) Anomalous sample closest to medioid }
	\label{trouser}
\end{figure}   	

On the other hand, in \autoref{trouser}, the samples containing trousers result in clear, unmistakable outliers. The medioid on the left appears to have stable topological structure, which is reflected by the a smaller number of in-set outliers, though as we can see in the central graph that they still exist. Additionally, the mean distance between the medioid and the anomaly set is significantly higher as this set introduces an entirely new connected component.

\subsubsection{Applications to other graph-structured data}
Additionally, the NAW metric can be demonstrated to be sufficiently general to apply to other instances of graph structured data. Here we test on a pair of widely used benchmark datasets for graph classification, PROTEINS and ENZYMES \cite{borgwardt2005shortest}. We selected these datasets due to the fact that they are vector attributed and thus are not fully represented by adjacency based metrics.

We compute the NAW distance by first generating the intrinsic and extrinsic pairwise distances between the vector-valued node coordinates. We define our intrinsic distance as the shortest path over the graph and our extrinsic distances as the point-to-point distances between the sets of points corresponding to nodes. For these data sets, we assign a probability density of $1/n$ where $n$ is the number of nodes in the graph. We then follow a common procedure for testing graph kernels, where we calculate the pairwise NAW distances between the graphs and use the distance matrix to generate a kernel matrix where each element is defined as $e^{-\gamma d(x_i, x_j)}$. A SVM is then trained using this kernel, and the parameters are tuned through 10-fold cross-validation. We compare our results to several existing graph kernels as well as other Wasserstein variants using this same methodology, using the results reported by their respective authors. 

\begin{table}[h!]
	\centering
	\begin{tabular}{||c ||c c ||} 
		\hline
		Method & PROTEIN & ENZYMES    \\ [0.5ex] 
		\hline\hline
		WL \cite{vishwanathan2010graph} & 72.9 & 53.7  \\ 
		GK \cite{shervashidze2009efficient} & 62.28  & -\\
		RW \cite{gartner2003graph} & 74.22& -  \\
		SP \cite{borgwardt2005shortest} & 75.07 & -    \\
		WL-OA \cite{kriege2016valid} &76.4 & 59.9 \\ 
		PSCN \cite{niepert2016learning} & 75.00 & -   \\ 
		FGW \cite{vayer2018fused} & 76.0 & 66.3 \\
		\hline
		\textbf{NAW} &  75.65 &  70.83 \\ [1ex] 
		\hline
	\end{tabular}
	\caption{Classification accuracy of SVM using various graph kernels.}
	\label{table:2}
\end{table}
In these experiments, the NAW was able to compete with state-of-the-art graph classification methods despite it's more general nature as a technique adapted to disconnected graphs. This was particularly notable on ENZYMES, as NAW outperformed more specialized techniques.

\section{Conclusion} 
We have proposed a new approach for the comparison of Mapper graphs and introduced a Wasserstein distance variant particularly suited for this mode of analysis. We have demonstrated that our metric captures variation in Mapper graphs in an intuitive way through a series of synthetic examples that capture desirable metric properties. Additionally, we have demonstrated the utility of the metric by utilizing it both in the Mapper comparison task and on broader problems involving graph structured data with vector-attributed nodes.

We believe that this new approach can make Mapper more accessible to a wider variety of uses by providing a stronger measure of the empirical stability of the algorithm over specific data. This can assist practitioners in identifying model drift and allow users to be more confident in existing results. 

\bibliographystyle{unsrtnat}
\bibliography{mappermetrics}

\begin{thebibliography}{10}

\bibitem{akoglu2015graph}
Leman Akoglu, Hanghang Tong, and Danai Koutra.
\newblock Graph based anomaly detection and description: a survey.
\newblock {\em Data mining and knowledge discovery}, 29(3):626--688, 2015.

\bibitem{melis17submodtransport}
D.~{Alvarez-Melis}, T.~S. {Jaakkola}, and S.~{Jegelka}.
\newblock {Structured Optimal Transport}.
\newblock {\em ArXiv e-prints}, December 2017.
\newblock \href {http://arxiv.org/abs/1712.06199} {\path{arXiv:1712.06199}}.

\bibitem{arjovsky2017wasserstein}
Martin Arjovsky, Soumith Chintala, and L{\'e}on Bottou.
\newblock Wasserstein generative adversarial networks.
\newblock In {\em International Conference on Machine Learning}, pages
  214--223, 2017.

\bibitem{earthquakes}
R.~Bendick and R.~Bilham.
\newblock Do weak global stresses synchronize earthquakes?
\newblock {\em Geophysical Research Letters}, 44(16):8320--8327.
\newblock URL:
  \url{https://agupubs.onlinelibrary.wiley.com/doi/abs/10.1002/2017GL074934},
  \href
  {http://arxiv.org/abs/https://agupubs.onlinelibrary.wiley.com/doi/pdf/10.1002/2017GL074934}
  {\path{arXiv:https://agupubs.onlinelibrary.wiley.com/doi/pdf/10.1002/2017GL074934}},
  \href {http://dx.doi.org/10.1002/2017GL074934}
  {\path{doi:10.1002/2017GL074934}}.

\bibitem{netsimile}
Michele Berlingerio, Danai Koutra, Tina Eliassi{-}Rad, and Christos Faloutsos.
\newblock Netsimile: {A} scalable approach to size-independent network
  similarity.
\newblock {\em CoRR}, abs/1209.2684, 2012.
\newblock URL: \url{http://arxiv.org/abs/1209.2684}, \href
  {http://arxiv.org/abs/1209.2684} {\path{arXiv:1209.2684}}.

\bibitem{borgwardt2005shortest}
Karsten~M Borgwardt and Hans-Peter Kriegel.
\newblock Shortest-path kernels on graphs.
\newblock In {\em Data Mining, Fifth IEEE International Conference on}, pages
  8--pp. IEEE, 2005.

\bibitem{bunke2007spectral}
Horst Bunke, Peter~J Dickinson, Miro Kraetzl, and Walter~D Wallis.
\newblock {\em A graph-theoretic approach to enterprise network dynamics},
  volume~24.
\newblock Springer Science \& Business Media, 2007.

\bibitem{carlsson2009topology}
Gunnar Carlsson.
\newblock Topology and data.
\newblock {\em Bulletin of the American Mathematical Society}, 46(2):255--308,
  2009.

\bibitem{oudottuning}
Mathieu Carri{\`e}re, Bertrand Michel, and Steve Oudot.
\newblock Statistical analysis and parameter selection for mapper.
\newblock {\em Journal of Machine Learning Research}, 19(12):1--39, 2018.
\newblock URL: \url{http://jmlr.org/papers/v19/17-291.html}.

\bibitem{carriere2017structure}
Mathieu Carriere and Steve Oudot.
\newblock Structure and stability of the one-dimensional mapper.
\newblock {\em Foundations of Computational Mathematics}, pages 1--64, 2017.

\bibitem{chandola2009anomaly}
Varun Chandola, Arindam Banerjee, and Vipin Kumar.
\newblock Anomaly detection: A survey.
\newblock {\em ACM computing surveys (CSUR)}, 41(3):15, 2009.

\bibitem{Chowdhury2018gromov}
S.~{Chowdhury} and F.~{M{\'e}moli}.
\newblock {The Gromov-Wasserstein distance between networks and stable network
  invariants}.
\newblock {\em ArXiv e-prints}, August 2018.
\newblock \href {http://arxiv.org/abs/1808.04337} {\path{arXiv:1808.04337}}.

\bibitem{courty2017optimal}
Nicolas Courty, R{\'e}mi Flamary, Devis Tuia, and Alain Rakotomamonjy.
\newblock Optimal transport for domain adaptation.
\newblock {\em IEEE transactions on pattern analysis and machine intelligence},
  39(9):1853--1865, 2017.

\bibitem{cuturi2013sinkhorn}
Marco Cuturi.
\newblock Sinkhorn distances: Lightspeed computation of optimal transport.
\newblock In {\em Advances in neural information processing systems}, pages
  2292--2300, 2013.

\bibitem{dey2016multiscale}
Tamal~K Dey, Facundo M{\'e}moli, and Yusu Wang.
\newblock Multiscale mapper: Topological summarization via codomain covers.
\newblock In {\em Proceedings of the twenty-seventh annual acm-siam symposium
  on discrete algorithms}, pages 997--1013. SIAM, 2016.

\bibitem{doshi2018movie}
Pratik Doshi and Wlodek Zadrozny.
\newblock Movie genre detection using topological data analysis.
\newblock In {\em International Conference on Statistical Language and Speech
  Processing}, pages 117--128. Springer, 2018.

\bibitem{remotesensing}
Ludovic Duponchel.
\newblock Exploring hyperspectral imaging data sets with topological data
  analysis.
\newblock {\em Analytica Chimica Acta}, 1000:123 -- 131, 2018.
\newblock URL:
  \url{http://www.sciencedirect.com/science/article/pii/S0003267017313077},
  \href {http://dx.doi.org/https://doi.org/10.1016/j.aca.2017.11.029}
  {\path{doi:https://doi.org/10.1016/j.aca.2017.11.029}}.

\bibitem{edelsbrunner2008persistent}
Herbert Edelsbrunner and John Harer.
\newblock Persistent homology-a survey.
\newblock {\em Contemporary mathematics}, 453:257--282, 2008.

\bibitem{flamary2017pot}
R{'e}mi Flamary and Nicolas Courty.
\newblock Pot python optimal transport library, 2017.
\newblock URL: \url{https://github.com/rflamary/POT}.

\bibitem{gartner2003graph}
Thomas G{\"a}rtner, Peter Flach, and Stefan Wrobel.
\newblock On graph kernels: Hardness results and efficient alternatives.
\newblock In {\em Learning theory and kernel machines}, pages 129--143.
  Springer, 2003.

\bibitem{hendrikson2016gromov}
Reigo Hendrikson et~al.
\newblock {\em Using Gromov-Wasserstein distance to explore sets of networks}.
\newblock PhD thesis, 2016.

\bibitem{asthma}
Timothy~SC Hinks, Xiaoying Zhou, Karl~J Staples, Borislav~D Dimitrov, Alexander
  Manta, Tanya Petrossian, Pek~Y Lum, Caroline~G Smith, Jon~A Ward, Peter~H
  Howarth, et~al.
\newblock Innate and adaptive t cells in asthmatic patients: relationship to
  severity and disease mechanisms.
\newblock {\em Journal of Allergy and Clinical Immunology}, 136(2):323--333,
  2015.

\bibitem{koutra2013deltacon}
Danai Koutra, Joshua~T Vogelstein, and Christos Faloutsos.
\newblock Deltacon: A principled massive-graph similarity function.
\newblock In {\em Proceedings of the 2013 SIAM International Conference on Data
  Mining}, pages 162--170. SIAM, 2013.

\bibitem{kriege2016valid}
Nils~M Kriege, Pierre-Louis Giscard, and Richard Wilson.
\newblock On valid optimal assignment kernels and applications to graph
  classification.
\newblock In {\em Advances in Neural Information Processing Systems}, pages
  1623--1631, 2016.

\bibitem{kusner2015word}
Matt Kusner, Yu~Sun, Nicholas Kolkin, and Kilian Weinberger.
\newblock From word embeddings to document distances.
\newblock In {\em International Conference on Machine Learning}, pages
  957--966, 2015.

\bibitem{prec_onc}
Jin-Ku Lee, Zhaoqi Liu, Jason~K. Sa, Sang Shin, Jiguang Wang, Mykola Bordyuh,
  Hee~Jin Cho, Oliver Elliott, Timothy Chu, Seung~Won Choi, Daniel I.~S.
  Rosenbloom, In-Hee Lee, Yong~Jae Shin, Hyun~Ju Kang, Donggeon Kim, Sun~Young
  Kim, Moon-Hee Sim, Jusun Kim, Taehyang Lee, Yun~Jee Seo, Hyemi Shin, Mijeong
  Lee, Sung~Heon Kim, Yong-Jun Kwon, Jeong-Woo Oh, Minsuk Song, Misuk Kim,
  Doo-Sik Kong, Jung~Won Choi, Ho~Jun Seol, Jung-Il Lee, Seung~Tae Kim, Joon~Oh
  Park, Kyoung-Mee Kim, Sang-Yong Song, Jeong-Won Lee, Hee-Cheol Kim, Jeong~Eon
  Lee, Min~Gew Choi, Sung~Wook Seo, Young~Mog Shim, Jae~Ill Zo, Byong~Chang
  Jeong, Yeup Yoon, Gyu~Ha Ryu, Nayoung K.~D. Kim, Joon~Seol Bae, Woong-Yang
  Park, Jeongwu Lee, Roel G.~W. Verhaak, Antonio Iavarone, Jeeyun Lee, Raul
  Rabadan, and Do-Hyun Nam.
\newblock Pharmacogenomic landscape of patient-derived tumor cells informs
  precision oncology therapy.
\newblock {\em Nature Genetics}, 50(10):1399--1411, 2018.

\bibitem{diabetes}
Li~Li, Wei-Yi Cheng, Benjamin~S. Glicksberg, Omri Gottesman, Ronald Tamler,
  Rong Chen, Erwin~P. Bottinger, and Joel~T. Dudley.
\newblock Identification of type 2 diabetes subgroups through topological
  analysis of patient similarity.
\newblock {\em Science Translational Medicine}, 7(311):311ra174--311ra174,
  2015.
\newblock URL: \url{http://stm.sciencemag.org/content/7/311/311ra174}, \href
  {http://arxiv.org/abs/http://stm.sciencemag.org/content/7/311/311ra174.full.pdf}
  {\path{arXiv:http://stm.sciencemag.org/content/7/311/311ra174.full.pdf}},
  \href {http://dx.doi.org/10.1126/scitranslmed.aaa9364}
  {\path{doi:10.1126/scitranslmed.aaa9364}}.

\bibitem{memoli2011gromov}
Facundo M{\'e}moli.
\newblock Gromov--wasserstein distances and the metric approach to object
  matching.
\newblock {\em Foundations of computational mathematics}, 11(4):417--487, 2011.

\bibitem{niepert2016learning}
Mathias Niepert, Mohamed Ahmed, and Konstantin Kutzkov.
\newblock Learning convolutional neural networks for graphs.
\newblock In {\em International conference on machine learning}, pages
  2014--2023, 2016.

\bibitem{orlin1993faster}
James~B Orlin.
\newblock A faster strongly polynomial minimum cost flow algorithm.
\newblock {\em Operations research}, 41(2):338--350, 1993.

\bibitem{osada2002shape}
Robert Osada, Thomas Funkhouser, Bernard Chazelle, and David Dobkin.
\newblock Shape distributions.
\newblock {\em ACM Transactions on Graphics (TOG)}, 21(4):807--832, 2002.

\bibitem{peyre2016gromov}
Gabriel Peyr{\'e}, Marco Cuturi, and Justin Solomon.
\newblock Gromov-wasserstein averaging of kernel and distance matrices.
\newblock In {\em International Conference on Machine Learning}, pages
  2664--2672, 2016.

\bibitem{ranshous2015anomaly}
Stephen Ranshous, Shitian Shen, Danai Koutra, Steve Harenberg, Christos
  Faloutsos, and Nagiza~F Samatova.
\newblock Anomaly detection in dynamic networks: a survey.
\newblock {\em Wiley Interdisciplinary Reviews: Computational Statistics},
  7(3):223--247, 2015.

\bibitem{RockWets98}
{R. Tyrrell} Rockafellar and Roger J.-B. Wets.
\newblock {\em Variational Analysis}.
\newblock Springer Verlag, Heidelberg, Berlin, New York, 1998.

\bibitem{rubner1998metric}
Yossi Rubner, Carlo Tomasi, and Leonidas~J Guibas.
\newblock A metric for distributions with applications to image databases.
\newblock In {\em Computer Vision, 1998. Sixth International Conference on},
  pages 59--66. IEEE, 1998.

\bibitem{brainmap}
Manish Saggar, Olaf Sporns, Javier Gonzalez-Castillo, Peter~A. Bandettini,
  Gunnar Carlsson, Gary Glover, and Allan~L. Reiss.
\newblock Towards a new approach to reveal dynamical organization of the brain
  using topological data analysis.
\newblock {\em Nature Communications}, 9(1):1399, 2018.

\bibitem{shervashidze2009efficient}
Nino Shervashidze, SVN Vishwanathan, Tobias Petri, Kurt Mehlhorn, and Karsten
  Borgwardt.
\newblock Efficient graphlet kernels for large graph comparison.
\newblock In {\em Artificial Intelligence and Statistics}, pages 488--495,
  2009.

\bibitem{shoubridge2002spectral}
Peter Shoubridge, Miro Kraetzl, WAL Wallis, and Horst Bunke.
\newblock Detection of abnormal change in a time series of graphs.
\newblock {\em Journal of Interconnection Networks}, 3(01n02):85--101, 2002.

\bibitem{singh2007topological}
Gurjeet Singh, Facundo M{\'e}moli, and Gunnar~E Carlsson.
\newblock Topological methods for the analysis of high dimensional data sets
  and 3d object recognition.
\newblock In {\em SPBG}, pages 91--100, 2007.

\bibitem{vayer2018fused}
T.~{Vayer}, L.~{Chapel}, R.~{Flamary}, R.~{Tavenard}, and N.~{Courty}.
\newblock {Optimal Transport for structured data}.
\newblock {\em ArXiv e-prints}, May 2018.
\newblock \href {http://arxiv.org/abs/1805.09114} {\path{arXiv:1805.09114}}.

\bibitem{vishwanathan2010graph}
S~Vichy~N Vishwanathan, Nicol~N Schraudolph, Risi Kondor, and Karsten~M
  Borgwardt.
\newblock Graph kernels.
\newblock {\em Journal of Machine Learning Research}, 11(Apr):1201--1242, 2010.

\bibitem{fmnist}
Han Xiao, Kashif Rasul, and Roland Vollgraf.
\newblock Fashion-mnist: a novel image dataset for benchmarking machine
  learning algorithms.
\newblock {\em CoRR}, abs/1708.07747, 2017.
\newblock URL: \url{http://arxiv.org/abs/1708.07747}, \href
  {http://arxiv.org/abs/1708.07747} {\path{arXiv:1708.07747}}.

\bibitem{zomorodian2005computing}
Afra Zomorodian and Gunnar Carlsson.
\newblock Computing persistent homology.
\newblock {\em Discrete \& Computational Geometry}, 33(2):249--274, 2005.

\bibitem{zong2018deep}
Bo~Zong, Qi~Song, Martin~Renqiang Min, Wei Cheng, Cristian Lumezanu, Daeki Cho,
  and Haifeng Chen.
\newblock Deep autoencoding gaussian mixture model for unsupervised anomaly
  detection.
\newblock In {\em International Conference on Learning Representations}, 2018.
\newblock URL: \url{https://openreview.net/forum?id=BJJLHbb0-}.

\end{thebibliography}

\end{document}